\documentstyle[12pt]{article}

\begin{document}

\title{Comment on the quantum modes of the scalar field on 
$AdS_{d+1}$ spacetime}


\author{Ion I. Cot\u aescu\\ {\small \it The West University of Timi\c soara,}\\
       {\small \it V. P\^ arvan Ave. 4, RO-1900 Timi\c soara, Romania}}

\date{\today}

\maketitle

\begin{abstract}
The problem of the quantum modes of the scalar free field on anti-de Sitter 
backgrounds with an arbitrary number of space dimensions is considered.  
It is shown that this problem can be solved by using the same quantum numbers 
as those of the nonrelativistic oscillator and two parameters which give 
the energy quanta and respectively the ground state energy. This last one is 
known to be just the conformal dimension of the boundary field theory of the 
AdS/CFT conjecture. 
  
Pacs 04.62.+v
\end{abstract}
\

The recent interest in propagation of  quantum scalar fields on anti-de 
Sitter (AdS) spacetime is due to the discovery of the AdS/Conformal field 
theory-correspondence \cite{M}. One of central points here is the relation 
between the field theory on the $(d+1)$-dimensional AdS ($AdS_{d+1}$) 
spacetime and the  conformal field theory on its $d$-dimensional 
Minkowski-like boundary ($M_{d}$). There are serious arguments that the local 
operators of the conformal field theory on $M_{d}$ correspond to the quantum 
modes of the scalar field on $AdS_{d+1}$ \cite{W}. Actually, for $d=3$ 
\cite{AIS,BF} as well as for any $d$ \cite{BL} it is proved that the conformal 
dimension in boundary field theory is equal with the ground state energy on 
$AdS_{d+1}$ \cite{W}. Moreover, it is known that the energy spectrum is 
discrete and equidistant \cite{BL} its quanta wavelength being just the 
hyperboloid radius of $AdS_{d+1}$. 

In these conditions  the scalar field on $AdS_{d+1}$  can be seen as the 
relativistic correspondent of the  nonrelativistic harmonic oscillator in $d$ 
space dimensions. This means that the  radial motion of the relativistic field 
may be governed by the same quantum numbers as that of the nonrelativistic 
oscillator, namely the radial and the angular quantum numbers. In the case of 
$d=3$ we know  that this is true \cite{AIS,N} but for $d>3$ the definition and 
the role of the angular quantum number are not completely elucidated. 

This is the motive why we would like to comment on this subject. Our aim is to 
present here the  form of the normalized wave functions of the scalar field on 
$AdS_{d+1}$ in terms of the above mentioned quantum numbers  and to establish 
the formula of the degree of degeneracy of the energy levels.
This problem is not  complicated but in arbitrary dimensions  some 
interesting technical details are worth reviewing. For this reason, we start 
with the separation of spherical variables in the Klein-Gordon equation on 
any central chart with $d$ space coordinates and then turn to the $AdS_{d+1}$ 
problem.    

Let us consider a static local chart of a $(d+1)$-dimensional spacetime where  
the  coordinates  $x^{\mu}$, $\mu=0,1, ..., d$ are the time,   $x^{0}=t$, 
and the Cartesian space coordinates  ${\bf x}\equiv (x^{1},\,x^{2},...,\,
x^{d})$, while the signature of the metric tensor $g_{\mu\nu}({\bf x})$ is 
$(+,-,-,...,-)$. The  charged scalar quantum field, $\phi\not=\phi^{+}$, of 
mass $M$, minimally coupled with the gravitational field, 
obeys the Klein-Gordon equation 
\begin{equation}\label{(kg)}
\frac{1}{\sqrt{g}}\partial_{\mu}\left(\sqrt{g}g^{\mu\nu}\partial_{\nu}\phi
\right) + M^{2}\phi=0\,, \quad
g=|\det(g_{\mu\nu})|\,, 
\end{equation} 
written in natural units with $\hbar=c=1$.
Since the chart is static there is a conserved  energy, $E$, and, 
consequently,  Eq.(\ref{(kg)}) has  particular solutions (of positive and 
negative frequency) of the form
\begin{equation}\label{(sol)}
\phi_{E}^{(+)}(t,{\bf x})=\frac{1}{\sqrt{2E}}e^{-iEt}U_{E}({\bf x}), \quad 
\phi^{(-)}=(\phi^{(+)})^{*},
\end{equation}
which give us the one-particle quantum modes.
These solutions  may be even square integrable functions or tempered 
distributions on the domain $D$ of the  space coordinates of the local chart. 
In  both cases they must be 
orhonormal (in usual or generalized sense) with respect to the relativistic 
scalar product \cite{BD} 
\begin{equation}\label{(sp1)}
\left<\phi_{E},\phi_{E'}\right>=i\int_{D}d^{d}x\,\sqrt{g}g^{00}\,
{\phi_{E}}^{*}\stackrel{\leftrightarrow}{\partial_{0}} \phi_{E'}
=\int_{D}d^{d}x\,\sqrt{g}g^{00}\, U_{E}^{*} U_{E'},
\end{equation}
which reduces to that of the static wave functions $U_{E}({\bf x})$.

In the following, we take into account only static central backgrounds that 
have static and spherically symmetric local charts where 
the line element is invariant under global rotations, $R\in SO(d)$, of 
the  Cartesian coordinates,  ${\bf x}\to {\bf x}'= R{\bf x}$. In these 
charts the metric is diagonal in  generalized spherical coordinates, 
$r, \theta_{1},...,\theta_{d-1}$, defined  as \cite{T}  
\begin{eqnarray}
x^{1}&=& r \, \sin\theta_{1}... \sin\theta_{d-1},
\nonumber\\ 
x^{2}&=& r \, \sin\theta_{1}... \cos\theta_{d-1},\label{(sfcor)}\\ 
\vdots\,\,\,&&  \nonumber\\
x^{d}&=& r \, \cos\theta_{1}\,.\nonumber
\end{eqnarray} 
such that the radial coordinate $r=|{\bf x}|$ be just the Euclidian norm of 
${\bf x}$. These coordinates cover the domain $D=D_{r}\times S^{d-1}$, i.e. 
$r\in D_{r}$ while ${\bf x}/r$ is on the sphere $S^{d-1}$. In general, the 
line element,   
\begin{equation}
ds^{2}=g_{\mu\nu}dx^{\mu}dx^{\nu}=g_{00}(r)dt^{2}+g_{rr}(r)dr^{2} +
g_{\theta\theta}(r)\, d\theta^{2}\,,
\end{equation}
with the angular part
\begin{equation}
d\theta^{2}={d\theta_{1}}^{2}+\sin^{2}\theta_{1}{d\theta_{2}}^{2}\, ...\, 
+\sin^{2}\theta_{1}\,\sin^{2}\theta_{2}...
\sin^{2}\theta_{d-2}{d\theta_{d-1}}^{2} 
\end{equation}
depends on three arbitrary functions of $r$,  $g_{00}$, $g_{rr}$ and 
$g_{\theta\theta}\equiv g_{\theta_{1}\theta_{1}}$.
Hereby we find that  
\begin{equation}
\sqrt{g({\bf x})}=\sqrt{\hat g(r)}\,(\sin\theta_{1})^{d-2} 
(\sin\theta_{2})^{d-3}\,...\,\sin\theta_{d-2}\,,
\end{equation}
where
\begin{equation}
\hat g=|g_{00}g_{rr}{g_{\theta\theta}}^{d-1}|.
\end{equation}
Furthermore, with the help of the new function
\begin{equation}
\rho={\hat g}^{1/4}|g_{rr}|^{-1/2}\,,
\end{equation}
we obtain the static Klein-Gordon equation
\begin{equation}\label{(kg1)}
\left[ \partial_{r}^{2}+2\partial_{r}(\ln\rho)\partial_{r}+g_{rr}
g^{\theta\theta}\Delta_{S}+g_{rr}M^{2}\right] U_{E}=g_{rr}g^{00}E^{2}U_{E}
\end{equation}
which concentrates all the angular derivatives in the angular Laplace 
operator  $\Delta_{S}$ \cite{T}. We observe that this equation becomes an 
energy squared eigenvalue problem in a {\em special} holonomic frame defined 
such that $g_{rr}=-g_{00}$. This condition can be achieved anytime with the 
help of an appropriate transformation of the radial coordinate of the central 
chart.       

The spherical variables of Eq.(\ref{(kg1)}) can be separated by using  
generalized spherical harmonics, $Y^{d-1}_{l\,(\lambda)}({\bf x}/r)$. 
These are eigenfunction of the angular Lalpace operator \cite{T},  
\begin{equation}
-\Delta_{S}
Y^{d-1}_{l\,(\lambda)}({\bf x}/r) 
=l(l+d-2)\,Y^{d-1}_{l\,(\lambda)}({\bf x}/r), 
\end{equation} 
corresponding to eigenvalues depending on the {\em angular} quantum number 
$l$ which  take  the  values $0,1,2,....$ \cite{T}. The notation $(\lambda)$ 
stands for a collection of quantum numbers giving the multiplicity of these 
eigenvalues \cite{T}, 
\begin{equation}
\gamma_{l}= (2l+d-2)\frac{(l+d-3)!}{l!\,(d-2)!}.
\end{equation}
Starting with  particular solutions  of the form 
\begin{equation}\label{(udex)}
U_{E,l(\lambda)}({\bf x})=\frac{1}{\rho(r)} R_{E,l}(r)\,
Y^{d-1}_{l\,(\lambda)}({\bf x}/r), 
\end{equation}
after a few manipulation, we find  the radial equation in a special frame
\begin{equation}
\left(-\frac{d^2}{dr^2}+g_{rr}g^{\theta\theta}l(l+d-2)-g_{rr}M^{2}+
\frac{1}{\rho}\frac{d^{2}\rho}{dr^2}\right)R_{E,l}=E^{2}R_{E,l}
\end{equation}
and the radial scalar product
\begin{equation}\label{(scprod)}
\left<R_{E,l},R_{E',l}\right>=\int_{D_{r}}dr \, 
R_{E,l}(r)^{*} R_{E',l}(r)\,.
\end{equation} 
Here  we have considered  that the generalized spherical harmonics are 
normalysed to unity with respect to their own angular scalar product defined 
on the sphere $S^{d-1}$. Thus we obtain an independent radial problem  in a 
special frame where  the radial scalar product is of the simplest form. This 
is the starting point for finding  analytical solutions of the Klein-Gordon 
equation on concrete central backgrounds.

Let us consider now the problem of the scalar field on  $AdS_{d+1}$. This 
is a hyperboloid in the $(d+2)$-dimensional flat spacetime of 
coordinates $Z^{-1},\, Z^{0},\, Z^{1},\,...,\, Z^{d}$  and  metric
\begin{equation}
\eta_{AB}={\rm diag}(1,1,-1,...,-1)\,, \quad A,\,B=-1,0,1,...d\,.
\end{equation}
The hyperboloid equation reads
\begin{equation}
(Z^{-1})^{2}+(Z^{0})^{2}-(Z^{1})^{2} ... -(Z^{d})^{2}=R^{2}
\end{equation}
where $R=1/\omega$  is its radius. In a special frame the  coordinates 
(\ref{(sfcor)}) satisfy    
\begin{eqnarray}
Z^{-1}\!\!&=&\frac{1}{\omega}\sec \omega r \cos \omega t\nonumber\\
Z^{0}&=&\frac{1}{\omega}\sec \omega r \sin \omega t \label{(ttt)}\\
{\bf Z}&=&\frac{1}{\omega}\tan \omega r \, \frac{{\bf x}}{r},\nonumber\\ 
\end{eqnarray} 
giving the line element \cite{AIS,BL}
\begin{equation}\label{(adsm)}
ds^{2}=\eta_{AB}dZ^{A}dZ^{B}=\sec^{2}\omega r\left(dt^{2}-dr^{2} -
\frac{1}{\omega^2}\sin^{2}\omega r\, d\theta^{2}\right)\,.
\end{equation}
on the radial domain $D_{r}=[0,\, \pi/2\omega)$. From Eqs.(\ref{(ttt)}) it 
results that the time of $AdS_{d+1}$ must satisfy   
$t\in [-\pi/\omega,\pi/\omega)$. We remind that  $t\in (-\infty,\infty)$ 
defines the universal covering spacetime of $AdS_{d+1}$ ($CAdS_{d+1}$) 
\cite{AIS}. 

Now from (\ref{(adsm)}) we  identify the  components of the metric tensor 
and we find
\begin{equation}\label{(rhor)}
\rho(r)=\left(\frac{1}{\omega}\tan \omega r\right)^{\frac{d-1}{2}}\,.
\end{equation}  
With these ingredients and by using the notations 
$\epsilon=E/\omega$ and $\mu=M/\omega$ (i.e.
$\epsilon=E/\hbar\omega$ and $\mu=Mc^2/\hbar\omega$ in usual 
units), we obtain the radial equation 
\begin{equation}\label{(radeq)}
\left[-\frac{1}{\omega^2}\frac{d^2}{dr^2}+\frac{2s(2s-1)}{\sin^{2}\omega r}+
\frac{2p(2p-1)}{\cos^{2}\omega r}\right]R_{E,l}=\epsilon^{2}R_{E,l}
\end{equation}
where 
\begin{equation}\label{(eqps)}
2s(2s-1)=\left(l+\frac{d}{2}-1\right)^{2}-\frac{1}{4}\,, \quad 
2p(2p-1)=\mu^{2}+\frac{d^{2}-1}{4}\,.
\end{equation}
It is well-known that the solutions of Eq.(\ref{(radeq)})  can be expressed in 
terms of hypergeometric functions \cite{AS}, up to normalization factors, as 
\begin{equation}\label{(gsol)}
R_{E,l}(r)\sim \sin^{2s}\omega r\cos^{2p}\omega r
F\left(s+p-\frac{\epsilon}{2},
s+p+\frac{\epsilon}{2}, 2s+\frac{1}{2}, \sin^{2}\omega r\right)\,.
\end{equation}
These radial functions can have good physical meaning only as polynomials 
selected by a suitable quantization condition. This is because the above 
hypergeometric  functions are so strongly divergent for 
$\sin^{2}\omega r\to 1$  that $R_{E,l}$ can not be interpreted as tempered 
distributions corresponding to a continuous energy spectrum. Therefore, 
we introduce the radial quantum number $n_{r}$ \cite{BL} and impose  
\begin{equation}\label{(quant)}
\epsilon=2 (n_{r}+s+p)\,,\quad n_{r}=0,1,2,...
\end{equation}
In addition, we choose the  positive solutions of Eqs.(\ref{(eqps)}) in order 
to avoid singularities in $r=0$ and $r=\pi/2\omega$. These are
\begin{equation}
2s=l+\frac{d-1}{2}\,,\quad  
2p=k-\frac{d-1}{2}\,,  
\end{equation} 
where we have denoted by
\begin{equation}
k=\sqrt{\mu^{2}+\frac{d^2}{4}}+ \frac{d}{2}  
\end{equation}
the conformal dimension of the field theory on $M_{d}$ \cite{W}. We note that 
(\ref{(quant)}) is the quantization condition on $CAdS_{d+1}$ while the 
$AdS_{d+1}$ one requires, in addition,  $k$ to be an integer number too 
\cite{AIS}.

The last step is to define the {\em main} quantum number, $n=2n_{r}+l$, which 
take the  values, $0,1,2,...$,  giving the energy levels 
\begin{equation}
E_{n}=\omega(k+n)
\end{equation}
If $n$ is even then $l=0,2,4,...,n$ while for odd $n$ we have $l=1,3,5,...,n$. 
In both cases we can demonstrate that the degree of degeneracy of the level 
$E_{n}$ is  
\begin{equation}
\gamma_{n}=\sum_{l}\gamma_{l}=\frac{(n+d-1)!}{n!\,(d-1)!}\,.
\end{equation}
Now it is a simple exercise to express (\ref{(gsol)}) in terms of Jacobi 
polynomials and to normalize them to unity with respect to (\ref{(scprod)}). 
Then by using (\ref{(rhor)}) and (\ref{(udex)}) we restore the final form of 
the solutions (\ref{(sol)}), 
\begin{equation}
\phi^{(+)}_{n,l(\lambda)}(t,{\bf x})=N_{n,l}\sin^{l}\omega r\cos^{k}\omega r 
{P_{n_{r}}^{(l+\frac{d}{2}-1,\,k-\frac{d}{2})}}(\cos 2\omega r)
Y^{d-1}_{l(\lambda)}({\bf x}/r)\,e^{-iE_{n}t}, 
\end{equation}
where  
\begin{equation}
N_{n,l}=\omega^{\frac{d-1}{2}}{\left[\frac{n_{r}!\,\Gamma(n_{r}+k+l)}
{\Gamma (n_{r}+l+\frac{d}{2})\Gamma(n_{r}+k+1-\frac{d}{2})}\right]}
^{\frac{1}{2}}\,.
\end{equation}

Thus we have shown that the problem of the one-particle quantum modes of the 
scalar field  on  $CAdS_{d+1}$  can be solved by using the quantum numbers 
$n/n_{r},l,(\lambda)$ and  parameters with precise physical interpretation, 
i.e. the frequency of the energy quanta, $\omega=1/R$, and the conformal 
dimension, k, which gives the ground state energy. We must specify that these 
results coincide with those of Refs.\cite{AIS,BF,N} for $d=3$ but in the 
general case of any $d$ these are similar (up to notations) to those of 
Ref.\cite{BL} only for $l=0$ while for $l\not=0$ there are some differences. 

Finally we note that the parameter $k$ we use instead of $M$ could play an 
important role in the supersymmetry and  shape invariance of the radial 
problem as well as in the structure of the dynamic algebra. The argument is 
that the radial problems for arbitrary $d$ are of the same nature as that 
with $d=1$ for which we have recently shown that $k$ determines the shape of 
the relativistic potential and, in addition, represents the minimal weight of 
the irreducible representation of its $so(1,2)$ dynamic algebra \cite{C}.

\end{document}